# CONVECTIVE TURBULENCE IN A RAYLEIGH BENARD SYSTEM WITH AND WITHOUT ROTATION IN THE INFINITE PRANDTL NUMBER LIMIT

**Jayanta K Bhattacharjee**

Harishchandra Research Institute
Jhunsi, Allahabad 211019
INDIA

## ABSTRACT

Convective turbulence in a Rayleigh Benard system has shown a marked reluctance to exhibit clear scaling in the energy or entropy spectrum. The recent numerical simulation of Pandey, Verma and Mishra has shown significantly better evidence of scaling in the infinite Prandtl number limit. This prompted us to look at this limit analytically. We find that the inevitable presence of sweeping helps give a very good understanding of the results of Pandey et al. In the presence of rotation the Rayleigh number dependence of the Nusselt number shows a strong increase in slope in recent experiments and simulations at finite Prandtl number. In this infinite Prandtl number we find that this steepening does not occur for any rotation speed and our results satisfy the Doering–Constantin bound.

Convection occurs in a fluid layer in a gravitational field if a temperature gradient is set up such that the lower layers of the fluid are hotter than the upper layers. This causes the fluid to be top-heavy and consequently a convective flow occurs which causes the hot fluid to rise and the cold fluid to fall. In a laboratory set-up or a simulation this is achieved by considering a fluid enclosed between two horizontal parallel plates a distance d apart, with the lower plate maintained at a temperature higher than that of the upper plate (Rayleigh-Benard system). The convective flow would be driven by the buoyancy force which is proportional to $\alpha(\Delta T)gd^3$ where $\alpha$ is the volume expansion coefficient, $\Delta T$ the temperature difference between the two plates, $g$ the acceleration due to gravity and $d^3$ represents a typical volume element of the fluid. The resistance to this flow would be provided by the retarding force due to the kinematic shear viscosity $\nu$ and also by the thermal diffusivity $\lambda$ which would tend to neutralize the temperature difference and thus prevent convection. Consequently the onset of convection is determined by a dimensionless number (Rayleigh number $R$) which is defined as $R = \dfrac{\alpha(\Delta T)gd^3}{\lambda \nu}$. The Rayleigh number has to cross a certain threshold (1708 for realistic boundary conditions and large aspect ratio containers) before the buoyancy overcomes the resistive effects and convection sets in as a steady cellular flow pattern. If the Rayleigh number is increased, the steady flow is destabilized by a time-dependent periodic flow and with increase of Rayleigh number the space and time dependence of the flow becomes more and more complicated leading to an essentially random flow pattern which is described as turbulent convection.

A measure of how strong the convective effects are is determined by the Nusselt number $Nu$ which is defined as the ratio of the total heat transferred from bottom to top to the total heat transferred due to convection. It follows that $Nu$ is also equal to unity added to the ratio of the heat transferred by convection to the heat transferred by conduction. Near the onset of convection, $Nu$ is slightly larger than unity and as $R$ is increased it becomes bigger and in the region of convective turbulence $Nu \gg 1$. In this range $Nu$ increases with $R$ and this increase can in general be expressed as a power law [1-7]. In many realistic systems (e.g. atmospheric flows) the convection occurs in a rotating fluid and hence it is of interest to consider a laboratory (or simulation) setup where the fluid container rotates with angular velocity $\Omega$ about the vertical (direction of gravity) axis [8-11]. The rotation can lead to instabilities in a viscous fluid under certain conditions and the effectiveness of rotation is consequently measured in terms of a dimensionless rotation rate called the Taylor number defined by $Ta = \dfrac{4\Omega^2 d^4}{\nu^2}$. Another important fluid parameter is the Prandtl number $\sigma = \nu/\lambda$ which compares the two diffusive processes of velocity and heat. A very large value of $\sigma$ would imply that at long time scales, the heat diffusion alone would be important while a small value of $\sigma$ corresponds to the long time importance of the momentum diffusion process.

If the turbulence is homogeneous and isotropic and there is no question of heat flow, then the most important paradigm is Kolmogorov's theory of turbulence [12] which asserts that the picture of steady turbulence involves a dissipation of kinetic energy due to viscosity at very small length scales (viscosity is a molecular process) and the injection of equal energy per unit time at a large scale which is of the order of the system size. The common rate of dissipation and injection is ε. In the intervening range of length scales (wave number scales in Fourier space) the energy is transferred from one scale to another (this is done by the nonlinear terms in the governing equations) at the same rate ε as discussed above. The intermediate range, called the inertial range, is far removed from the scales where injection and dissipation occur and hence is a universal range for all turbulent fluids where the energy spectrum $E_v(k)$ is determined by $k$ and $ε$ alone. The total energy $E_v$ per unit mass is written as $E_v = \int E_v(k) dk$ to define the energy spectrum. According to Kolmogorov, $E_v(k) \propto k^{-5/3}$.

When one attempts to derive this dimensional analysis result from the Navier Stokes equation, one encounters a problem. An approximate picture for the Fourier transform $\tilde{v}(k)$ of the velocity is that of an eddy. The Navier Stokes equation couples all pairs of $\tilde{v}(p)$ and $\tilde{v}(q)$ with $p$ and $q$ adding upto $k$. The typical size of an eddy is $1/k$. When a large eddy ($p \approx 0$) couples with a small eddy ($q \approx k$) we have the dynamics of a small eddy simply being transported by the large eddy. This advection is not turbulent and simply corresponds to a sweeping of small eddies by a large one [13]. The sweeping effect leads to $E_v(k) \propto k^{-3/2}$ and needs to be removed from the analysis of the turbulent spectrum to see clearly the Kolmogorov spectrum. Since 5/3 is not all that different from 3/2 the sweeping is difficult to detect.

When it comes to convective turbulence, the application of the Kolmogorov theory has to recognize that there is, in addition to the kinetic energy, another "energy" whose density is proportional to the square of the local fluctuation in the temperature. If we call this the thermal energy then this too has a flux from the large length scales to the small length scales. If the thermal flux dominates the kinetic energy flux then the energy spectrum $E_v(k)$ is determined by the thermal flux and a dimensional analysis gives $E_v(k) \propto k^{-11/5}$ which is called the Bolgiano-Obukhov spectrum [14-16]. Thus in convective turbulence we can have two kinds of spectrum, one proportional to $k^{-5/3}$ and other proportional to $k^{-11/5}$. The spectrum associated with the temperature fluctuation is generally called the entropy spectrum.

For Rayleigh numbers $R$ significantly higher than the threshold of convection, the Nusselt number ($Nu$) scales [1-6] with the Rayleigh number $R$ according to $Nu \propto R^\beta$. The exponent $β$ has a slow variation with $R$ [5]. In the lower range of $R$ values it is nearly 2/7 and increases slowly with $R$ presumably towards what is called the ultimate regime [7] of $β=1/2$. The effect of rotation on this convective turbulence has also been widely studied. A new dimension in these studies was introduced by Schmitz and Tilgner [8] and by King et. al. [9] when they observed that for the rotating Rayleigh Benard system the scaling of $Nu$ with $R$ follows a significantly larger exponent for the lower range of $R$ values. How low is the lower range depends on both the rotation rate and the Prandtl number( the ratio of kinematic viscosity to the thermal diffusivity).This increased value of the exponent $β$ is also dependent on the Prandtl number and it was found [8,9] that for a Prandtl number of 7 the exponent was about 1.2 while for a Prandtl number of 0.1 it varied [10] between 0.5 and 0.7. However, it was established by Doering and Constantin [11] that in the limit of the Prandtl number being infinity there is a strict upper bound on the Nusselt number independent of the rotation rate and this bound is given by $c R^{2/5}$ where $c$ is a numerical constant depending on the rotation rate. This motivates a direct calculation of the Nusselt number in the infinite Prandtl number limit.

Turning to the energy and entropy spectra, the conventional wisdom for zero rotation rate has been that the kinetic energy spectrum will be Kolmogorov [12] like ($k^{-5/3}$) for large values of $k$ (small length scales) and Bolgiano-Obukhov [14-16] like ($k^{-11/5}$) for small values of $k$ (large length scales), $k$ being the wave number. The dividing length scale of the two spectra is the so-called Bolgiano length which can be expressed in terms of the Nusselt, Rayleigh and Prandtl numbers. Intensive numerical and experimental work [17-19] over the last decade has not seen any convincing scaling behaviour in this problem. However, in the infinite Prandtl number limit the very recent work of Pandey, Verma and Mishra [20] show a significant scaling regime with an exponent that is expected from dimensional arguments. However, there is a small systematic deviation from pure scaling making an investigation of this infinite Prandtl number limit worthwhile. The effect of rotation on the spectra has not been studied at all giving further incentive for a direct calculation.

In view of the above we decided to explore the infinite Prandtl number limit by diagrammatic perturbation theory arranged by the number of fully dressed loops [21,22]. The scaling forms are independent of the order of perturbation theory while the amplitudes associated with the scaling will be dependent on the order of perturbation theory. We find that

A) the kinetic energy spectrum at zero rotation speed is a combination of $k^{-13/3}$ and $k^{-7/2}$, the latter coming from the sweeping effect and being the cause of the departure from pure scaling as observed by Pandey et al. [20].

B) at very high rotation speeds the spectrum scales as $k^{-7}/\Omega^{4/3}$, where $\Omega$ is the rotation rate

C) for zero rotation rate, $Nu \propto R^{1/3}$

D) for very high rotation rates, $Nu \propto R^{1/3}/\Omega^{2/3}$.

We begin by writing down Navier Stokes equation and the heat diffusion equation for a rotating fluid in the Boussinesque approximation (this is an approximation in which, after the buoyancy force is accounted for, the fluid is treated as incompressible). The variables that are used are the ones which describe the deviation from the conduction state described by $\vec{v} = 0$ and a temperature profile of $T_s(z) = T_1 - \dfrac{\Delta T}{d} z$, where $\Delta T = T_1 - T_2$ and $z$ is the height from the bottom plate. The fluid is taken to be confined between two large horizontal parallel plates placed at $z=0$ and $z=d$ with the lower plate maintained at a temperature $T_1$ and the top plate maintained at $T_2$ with $T_1 > T_2$. The system rotates with an angular velocity $\Omega$ about the vertical ($z$) axis. The fluid has kinematic viscosity $v$ and thermal diffusivity $\lambda$, so that the Prandtl number $\sigma$ is given by $\sigma=v/\lambda$. The equations of motion will be written in terms of dimensionless variables. The velocity will be scaled by $\lambda/d$, time by $d^2/\lambda$, and the temperature deviation from the steady conduction state will be scaled by $\Delta T$ and denoted by $\theta$. After the divergence free condition on the velocity field is used to eliminate the pressure fluctuation, one gets the standard form

(1a) $\quad \dfrac{1}{\sigma}\nabla^2(\dot{u}_\alpha + (u_\beta \partial_\beta)u_\alpha) = R(\nabla^2 \delta_{\alpha 3} - \partial_\alpha \partial_3)\theta + \nabla^2 u_\alpha + Ta^{1/2}(\nabla^2 \varepsilon_{\alpha\beta 3} u_\beta - \varepsilon_{\gamma\beta 3}\left(\dfrac{\partial^2 u_\beta}{\partial x_\alpha \partial x_\gamma}\right))$

(1b) $\quad\quad\quad\quad\quad\quad\quad \dot{\theta} + (u_\alpha \partial_\alpha)\theta = \nabla^2 \theta + u_3 + f$

In the above $R$ is the Rayleigh number, $Ta$ is the Taylor number, and $f$ is a random external force which will help define correlation functions and propagators subsequently. The random force $f$ is Markovian and has a Gaussian distribution with the odd moments vanishing and the second moment given by (in Fourier space)

(2) $\quad\quad\quad\quad\quad\quad\quad < f(k,\omega) f(k',\omega') > = D(k)\delta(\omega+\omega')$

The function $D(k)$ will be dictated by the physical requirements of the theory. In the limit of $\sigma \to \infty$, the terms on the left hand side of Eq. (1a) disappear and the velocity field is tied to the temperature field by the relation (in wave-number space)

(3) $\quad\quad\quad\quad (k^6 + Tak_3^2)u_\alpha(k) = RP_{\alpha 3}(k)k^4 \theta(k) - Ta^{1/2}\varepsilon_{\alpha\beta 3}k_\beta k_3 \theta(k)$

where $P_{\alpha\beta}(k) = \delta_{\alpha\beta} - \dfrac{k_\alpha k_\beta}{k^2}$ is the usual projection operator. It should be noted that R is still finite as it contains three independent quantities $\Delta T$, $\nu$ and $\lambda$. Using Eq. (3) we can write Eq. (1b) in the form:

(4) $\quad\quad\quad\quad \dot{\theta}(k) + \Gamma(k)\theta(k) = -ik_\alpha \int \dfrac{d^3 p}{(2\pi)^3} u_\alpha(\vec{p})\theta(\vec{k}-\vec{p}) + f(k)$

where $\Gamma(k)$ is the relaxation rate of a fluctuation at wave number $k$. Clearly $\Gamma(k) \propto k^2$ (for $k$ greater than a minimum value of $O(R^{1/4})$) to the leading order with corrections, coming from the $R$-dependent term, which are not isotropic in $k$-space. At present we do not bother with this complication as it is the nonlinear contribution to relaxation that will dominate.

We will treat the nonlinear term in Eq. (3) in perturbation theory. There are two primary ingredients in the perturbation theory- the Green's function $G(k,\omega)$ defined as $<\dfrac{\delta\theta(k,\omega)}{\delta f(k',\omega')}> = G(k,\omega)\delta(\vec{k}+\vec{k}')\delta(\omega+\omega')$ and the correlation function $C(k,\omega)$ defined as $<\theta(\vec{k},\omega)\theta(\vec{k}',\omega')> = C(k,\omega)\delta(\vec{k}+\vec{k}')\delta(\omega+\omega')$ where the angular brackets denote averaging over the random force $f$. The linear theory gives the Green's function $G_0(k,\omega) = (-i\omega + \Gamma k^2)^{-1}$, while the full Greens function is given by Dyson's equation

(5) $$G^{-1}(k,\omega) = G_0^{-1} + \Sigma(k,\omega) = -i\omega + \Gamma k^2 + \Sigma(k,\omega)$$

where $\Sigma(k,\omega)$ is the self energy which is to be interpreted as the contribution of the nonlinear terms to the relaxation rate.

Our interest will be in seeking scaling solutions $\Sigma(k,0) \propto k^n$ and $C(k) = \int \frac{d\omega}{2\pi} C(k,\omega) \propto k^{-m}$. The two quantities which are usually measured in experiments and numerical simulations are the spectrum $E_v(k)$ and the Nusselt number $Nu(R)$. The spectrum of the temperature variable $\theta(k)$ is related to the correlation function $C(k)$ in a 3-dimensional space by the relation

(6) $$E_v(k) = C(k)k^2 \propto k^{2-m}$$

It should be clearly understood that in the above equation we are looking at the isotropic part of $C(k)$ alone. As our calculations will clearly show there will be strong anisotropy in the correlation function as explored in great details in the numerical work of Rincon [23]. The kinetic energy spectrum which is the one which is usually more popular can be found trivially from Eq. (3) once $C(k)$ is determined. The other measured quantity, the Nusselt number, is expressed in terms of a volume averaged correlation function as [using Eq. (3)]

(7) $$Nu = \frac{1}{V}\int d^3r <u_3(\vec{r},t)\theta(\vec{r},t)> = R\int \frac{d^3p}{(2\pi)^3} C(p) p^2 \frac{\sin^2\gamma}{p^4 + Ta\cos^2\gamma}$$

where $\gamma$ is the angle between the vector **p** and the vertical so that $d^3p = 2\pi p^2 \sin\gamma\, d\gamma\, dp$ Clearly once $C(k)$ is known both the usually measured quantities can be determined. We now argue that for all practical purposes $C(k)$ can be found from a knowledge of the self energy $\Sigma(k,\omega)$. The turbulent regime is dominated by the nonlinear terms and hence in this regime the self energy will exceed the bare relaxation rate $\Gamma(k)$. Further since our interest will be in long time properties, we will explore only the low frequency regime and hence the frequency dependence of $\Sigma(k,\omega)$ will not play any significant role. We will therefore incorporate the effect of the nonlinear term in the dynamics of Eq. (4) through a relaxation rate $\Sigma(k)$. This allows us to write Eq. (4) in frequency space, through an equivalent linearization, as $(-i\omega + \Sigma(k))\theta(k,\omega) = f(k,\omega)$ and it follows that

(8) $$C(k) = \int \frac{d\omega}{2\pi} <\theta(\vec{k},\omega)\theta(-\vec{k},-\omega)> = \int \frac{d\omega}{2\pi} \frac{<ff>}{\omega^2 + \Sigma(k)^2} = \frac{D(k)}{\Sigma(k)}$$

Thus we see that $\Sigma(k)$ plays the pivotal part in the calculations. The form of $D(k)$ will be forced on us by the requirement of a constant flux of the $\theta$ related "energy" through wave-number space.

Substituting for $u(k)$ in Eq. (4) from Eq. (3) wehave a nonlinear equation in the single field $\theta(k,t)$ and one can use the usual rules for diagrammatic perturbation theory [21] to calculate $\Sigma(k)$, which is the primary aim of the theory. In the fully self-consistent one loop [22] approximation, straightforward algebra leads to the answer:

(9)
$$\Sigma(k) = \int \frac{d^3 p}{(2\pi)^3} \left\{ \frac{[R^2(k_3 p^2 - p_3(\vec{k}.\vec{p}))^2 + Tap^2 p_3^2 (k_1^2 p_2^2 + k_2^2 p_1^2)]}{p^4 [\Sigma(p) + \Sigma(|\vec{k}-\vec{p}|)]} \right\}$$
$$\times \left\{ \frac{C(p) p^8}{(p^6 + Tap_3^2)^2} + \frac{C(|\vec{k}-\vec{p}|)|\vec{k}-\vec{p}|^8}{(|\vec{k}-\vec{p}|^6 + Ta(k_3 - p_3)^2)^2} \right\}$$

This is one of the central equations of this paper which will be used repeatedly. Before we make use of this equation we need to introduce the other ingredient of the theory – the rate at which the "thermal" energy $\int \theta^2 d^3 r / 2$ changes. This is the rate at which energy is dissipated at short length scales and the rate at which it is injected by the external random force $f$ at large length scales. It is also the rate at which the nonlinear term in Eq. (4) transfers the energy from one wave-number to another. This transfer rate is $\Pi(k)$ and from Eqs. (4) and (3) can be written as the expectation value

(10)
$$\Pi(k) = -iR \int_0^k \frac{d^3 p}{(2\pi)^3} \int \frac{d^3 q}{(2\pi)^3} \frac{p_\alpha P_{\alpha 3}(q) q^4 < \theta(\vec{q}) \theta(\vec{p}-\vec{q}) \theta(-\vec{p}) >}{q^6 + Taq_3^2}$$

where we have ignored the anisotropic part of $\Pi(k)$ and focussed on the part which depends on the magnitude of $k$ alone. To evaluate the three point correlation function above it is useful to write Eq. (4) as the integral equation [we have replaced $u$ by $\theta$ using Eq. (3)]

(11)
$$\theta(\vec{k},t) = \theta_0(\vec{k},t) + iR \int_0^t dt' G(k,t-t') \int d^3 p \left\{ \frac{1}{(2\pi)^3} \frac{k_\alpha p^4 P_{\alpha 3}(p)}{p^6 + Tap_3^2} \theta(\vec{p},t') \theta(\vec{k}-\vec{p},t') \right\}$$

where $\theta_0(\vec{k},t)$ is the solution of Eq. (4) without the non-linear term. We can insert the above expression in Eq. (10) and iteratively evaluate $\Pi(k)$.

Having introduced the two primary quantities, $\Sigma(k)$ and $\Pi(k)$, we are ready to look at the consequences. We focus on the non rotating case first i.e. set $\Omega=0$. In this case Eq. (9) becomes

(12)
$$\Sigma(k) = R^2 \int \frac{d^3 p}{(2\pi)^3} \frac{(k_3 - \frac{(\vec{k}.\vec{p})}{p^2} p_3)^2}{\Sigma(p) + \Sigma(|\vec{k}-\vec{p}|)} \left[ \frac{C(p)}{p^4} + \frac{C(|\vec{k}-\vec{p}|)}{|\vec{k}-\vec{p}|^4} \right]$$

It is clear that the two terms in the square bracket have two different characteristics. The first clearly has a divergence as $p \to 0$ and would require the integral to be cut off at some lower limit. In this case we can easily read off that $\Sigma(k)$ is proportional to $k$. This means that the exponent '$n$' that we had introduced would be given by $n=1$. This is the so called sweeping term. It corresponds to a $\theta$ - fluctuation at wave number $k$ being simply transported by a long wavelength velocity fluctuation and hence the relaxation rate is simply proportional to the wave-vector. The second term in the square bracket in Eq. (12) gives a finite contribution and hence the scaling characteristic of that part of the contribution to the relaxation rate yields on a simple power counting

(13) $$m+2n=1$$

We now turn to the flux $\Pi(k)$ at zero rotation rate. Looking at Eqs. (10) and (11) we see that the right hand side of Eq. (11) will contain one Greens function and two correlation functions. Since the Greens function is the inverse of the self energy, the power count of the relevant integral yields

(14) $$2m+n=4$$

for the flux to be a constant independent of $k$. From Eqs. (13) and (14) we find $n=-2/3$ and $m=7/3$. The corresponding kinetic energy correlation function [from Eq. (3)] is seen to scale as $k^{-19/3}$ and hence the energy spectrum will scale as $k^{-13/3}$. If the dynamics is sweeping dominated then Eq. (13) is irrelevant as $n$ gets fixed at 1 and for a constant flux [Eq. (14)] we get $m=3/2$. The kinetic energy spectrum is then given by $k^{-7/2}$. In the most simple minded handling of these two contributions to the spectrum, the result would be an additive form (consistent with a lowest order calculation of the correlation function)

(15) $$E_v(k) = \frac{c_1}{k^{13/3}} + \frac{c_2}{k^{7/2}}$$

where $c_1$ and $c_2$ are constants. The validity of Eq. (15) is adequately borne out by the data of Pandey, Verma and Mishra [20]. A spectrum compensated by $k^{13/3}$ has a small but distinct increasing trend with increasing $k$ while a spectrum compensated by $k^{7/2}$ has a decreasing trend. A complete scenario is shown in Fig 1 where the fit to the spectrum demonstrates convincingly that sweeping cannot be ignored.

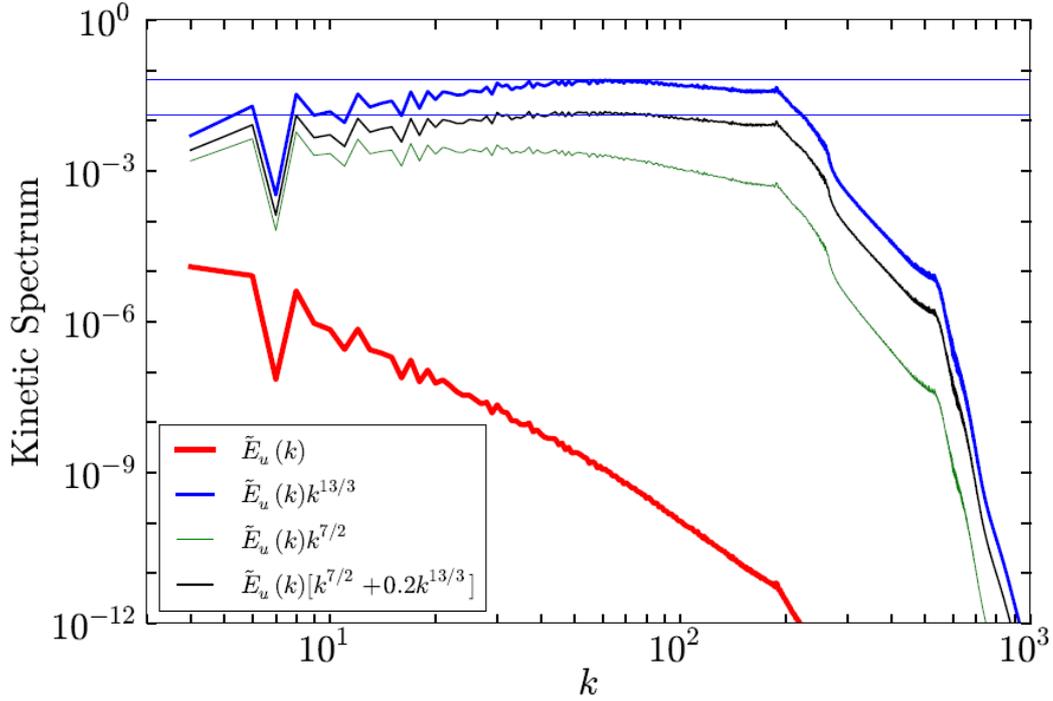

We now note that the result obtained for the case without sweeping ($m=7/3$, $n=-2/3$) actually fixes the form of the noise correlation function $D(k)$. Noting that $C(k)=D(k)/\Sigma(k)$ according to Eq. (8), we find that $D(k) \propto k^{-3}$. This form of $D(k)$ is completely consistent with that used by various authors [23-26] in arriving at the Kolmogorov spectrum. Recognizing this form of $D(k)$ and its relation to $C(k)$ will be very useful in discussing the finite rotation case. This is because $D(k)$ cannot be affected by Rayleigh number or Taylor number, having to do with an external random force, and the calculation shifts to the determination of $\Sigma(k)$ alone.

Turning to the Nusselt number for zero rotation rate, we see that we need to find the $R$ dependence of the amplitudes of $\Sigma(k)$ and $C(k)$. If we write $\Sigma(k) = \Gamma_0 k^n$ and $C(k) = C_0 k^{-m}$, then from Eq. (12) it is clear that $\Gamma_0^2 / C_0 R^2$ is a constant independent of $R$. In a similar fashion from Eqs. (10) and (11), we find the combination $R^2 C_0^2 / \Gamma_0$ is independent of $R$. This leads to $C_0 \propto R^{-2/3}$. From Eq. (7) it follows immediately that

(16) $$Nu \propto R^{1/3}$$

It should be clear that this result holds independent of whether one is discussing the sweeping free regime or the sweeping independent one.

We now turn to the case of finite rotation rate. Now we do not have any divergence at small wave numbers and the difference between the sweeping and non-sweeping regimes disappear. To find the effect of rotation on $\Sigma(k)$, it is simplest to talk about the very high

Taylor number regime when in the integrand of Eq. (9) we can drop the external wave number $k$. Since we are looking at the isotropic part of $\Sigma(k)$ alone we can drop the $Ta$ containing terms in the numerator of Eq. (9). We assume that for $Ta\gg 1$,

$$\Sigma(k) = \Gamma_1 \frac{k^{n'}}{Ta^y} \tag{17}$$

where $\Gamma_1$ is a constant. We insert this form in Eq. (9), remember Eq. (8) and by simple power counting arrive at

$$\Sigma(k) = const. \frac{k^2}{Ta^{1+(n'/2)-2y}} \tag{18}$$

Comparing Eqs (17) and (18), we conclude that $n'=2$ and $y=2/3$. In this limit, we immediately have $C(k) \propto Ta^{2/3}/k^5$ and hence the kinetic energy spectrum will scale as

$$E_v(k) = c Ta^{2/3}/k^7 \tag{19}$$

where $c$ is a constant. We return to Eq. (7) and note that $C(k)$ is still proportional to $R^{-2/3}$ and hence the $C(p)$ appearing in Eq. (7) will be proportional to $Ta^{2/3}R^{-2/3}p^{-5}$ giving the result

$$Nu \propto R^{1/3}/Ta^{1/3} \propto R^{1/3}/\Omega^{2/3} \tag{20}$$

Thus we find that in this infinite Prandtl number limit, the $R$ dependence is always $R^{1/3}$ with the prefactor being a function of $\Omega$, the rotation rate. This result supports the strict upper bound on the Nusselt number found by Doering and Constantin [11].

To get a feel for the appropriateness of the claim that in this limit the Nusselt number has just one power law dependence on $R$, one can have a look at the numerical simulations of King et al. One can see that as the Prandtl number is increased (Fig 2b of King et al [9]) the steepening range becomes smaller and smaller and is hardly visible at a Prandtl number of 100.

In conclusion we have shown that in this infinite Prandtl number limit the explicit calculation with Eulerian hydrodynamics brings out clearly the role of the sweeping and gives the self energy explicitly as the sum of a sweeping and a non-sweeping term. This result is supported very strongly by the numerical work of Pandey et al. [20]. We also predict that at high rotation speeds the spectrum will be much steeper and have a characteristic rotation rate dependence. The Nusselt number calculation gives a Rayleigh number dependence in accordance with the Doering–Constantin [11] bound and a characteristic dependence on rotation speed at high Taylor numbers.

I am grateful to Ambrish Pandey and Mahendra K Verma for several interesting discussions and for providing Fig. 1 of this paper from their unpublished data.

FIGURE CAPTIONS

Fig 1. The kinetic energy spectrum as a function of the wave-number *k*. The compensated spectrum with the sweeping contribution included is the best fit to the data. The contribution of sweeping is quite strong as seen from the coefficients of the two terms.